\documentclass[a4paper,twoside]{article}

\usepackage{epsfig}
\usepackage{subfigure}
\usepackage{calc}
\usepackage{amssymb}
\usepackage{amstext}
\usepackage{amsmath}
\usepackage{amsthm}
\usepackage{multicol}
\usepackage{pslatex}
\usepackage{SCITEPRESS}     

\begin{document}

\title{Toward a Better Understanding of How to Develop Software Under Stress -- Drafting the Lines for Future Research}

\author{\authorname{Joseph Alexander Brown, Vladimir Ivanov, Alan Rogers, \\ Giancarlo Succi, Alexander Tormasov, and Jooyong Yi}
\affiliation{Innopolis University, Universitetskaya St, 1, Innopolis, Russia, 420500}
\email{\{j.brown,v.ivanov,a.rogers,g.succi,tor,j.yi\}@innopolis.ru}
}

\keywords{Software development under adverse circumstances, empirical software engineering, software quality.}

\abstract{Software is often produced under significant time constraints. Our idea is to understand the effects of various software development practices on the performance of developers working in stressful environments, and identify the best operating conditions for software developed under stressful conditions collecting data through questionnaires, non-invasive software measurement tools that can collect measurable data about software engineers and the software they develop, without intervening their activities, and biophysical sensors and then try to recreated also in different processes or key development practices such conditions.
}

\onecolumn \maketitle \normalsize \vfill

\section{\uppercase{Introduction}}
\label{S.Introduction}
Software is often produced under significant time constraints. Our idea is to understand the effects of various software development practices on the performance of developers working in stressful environments, and identify the best operating conditions for software developed under stressful conditions. To achieve this goal, we argue to divide the research in the following two phases: ``in vitro'' and ``in vivo''. 

In the ``in vitro'' phase, the conditions under which people operate the best will be identified and monitored by collecting data through questionnaires, non-invasive software measurement tools that can collect measurable data about software engineers and the software they develop, without intervening their activities, and biophysical sensors.

In the ``in vivo'' phase,  the best working conditions identified in the earlier ``in vitro'' phase will be recreated in order to study their effects in various stressful conditions. In this phase, it will also be investigated the effects of well-known development practices such as pair programming, test driven development, inspection, collective code ownership, constant integration.

In the next section we briefly survey the state of the art and related works. Then, in Section \ref{problem} we define the problem statement and specific research questions, Finally, in Section \ref{approach} we present our view of the possible solution of the problem and concrete approach to the novel research agenda.

\section{\uppercase{Related Works}}

\subsection{Software process improvement}
The work on software process improvement has spanned decades using various methodologies \cite{Succi:C2.1989,Succi:J20.1997,Succi:C82.2000}, processes \cite{Succi:C73.2000,Succi:C224.2010,Succi:J83.2012,Succi:C245.2013,Succi:C144.2004}, and devices \cite{Succi:C236.2011,Succi:C244.2013} and there is a large corpus of scientific studies referring to it as it is evidenced by the recent literature reviews on the subject
\cite{khan2017systematic}. 
The discipline is now moving to acknowledge specific aspects of it, like SMEs working on web-based systems \cite{sulayman2009systematic}, 
process and simulations \cite{ali2014systematic}, 
agile methods \cite{campanelli2015agile}. 

Particular relevance is now placed on empirical evaluations of new approaches \cite{unterkalmsteiner2012evaluation,Succi:J100.2015}. 
The proposed work moves exactly along these lines, proposing new approaches to a particularly difficult development process centered on a clear empirical understanding of the best conditions under which software developers and engineers produce their work.

\subsection{Influence of the state of mind on the quality of developers}

It is a well known fact that the state of mind influences work and that especially positive feeling tend to be correlated with high quality work, especially in knowledge-intensive fields, as discussed in multiple research works, like the one of \cite{Amabile-Harvard96,Succi:J60.2004,lyubomirsky2005benefits,barsade2007does,Baas-Psychological08,Succi:c244.2012,Succi:J91.2013} 

The influence of the state of mind on the quality of the software being developed has been recognized since the early stages of software engineering. From the 1950-s there have been studies trying to understand the psychological profiles of developers acknowledging the intrinsic connection that exists between the state of the mind and the quality of the code, like the work of Rowan \cite{rowan1957psychological}, 
and the role of personalities and of interpersonal communications has been a central part of the agile approaches to software development as championed by the works of Cockburn and Highsmith \cite{cockburn2001agile}, Williams and Cockburn \cite{DBLP:journals/computer/WilliamsC03,Succi:C217.2009,feldt2010links,denning2012moods}, 
and others.

There has also been a significant literature evidencing that happiness and positive feelings have a positive impact on quality and productivity in the workplace and specifically promotes creativity \cite{brand2007we,davis2009understanding}. 
This is particularly important in software development, which includes a high amount of creativity as it has been acknowledged for many years now in several research work like \cite{fischer1987cognitive,glass1992software,shaw2004emotions,Knobelsdorf:2008:CPC:1597849.1384347,lewis2011affective}.

More recently, there have been studies linking the specific concept of well-being and concentration to the effectiveness in producing quality software. In 2002, Succi et al. have conducted one of the first research endeavours linking specific software practices to job satisfaction and low turnover \cite{succi2002preliminary}, 
and then creating a model for explaining job satisfaction and its influence on quality and productivity \cite{pedrycz2011model}, 
exploring how developers move in their workplace \cite{DBLP:conf/tools/CorralSSSV12} 
and relating pair programming with developers attention \cite{sillitti2012understanding}.

Scientific research has also explored the specific concept of happiness at work, connecting it to high-quality software artifacts like the works of \cite{khan2011moods,graziotin2013happy,graziotin2014happy,murgia2014developers}.

In all these studies the main vehicle for collecting data have been questionnaires and subjective evaluations. Biophysical signals have not been used. Some research has already performed also using such signals using suitable devices, and it is concentrated in mainly three research units.

\subsection{Studies of biometric sensors to evaluate the state of the mind of developers}
In this subsection we concentrate the description on the key studies using biometrics sensors to evaluate the state of the mind of developers and their relationships with tasks to accomplish. There is not any significant research effort on how to develop software under stress. Fritz at al. obtain metrics that correlate with software developers performance. In \cite{zuger2015interruptibility} 
they used interruptibility while \cite{muller2015stuck} 
used positive and negative emotions of software developers as metrics of progress in the change task.

They analyze data from multiple bio-sensors, including eye trackers for measuring pupil size and eye blinks, electroencephalography to determine brain activity, electrodermal activity sensors to detect skin-related activity, and heart-related sensors. They apply methods of supervised learning (Naive Bayes) to distinguish levels of these cognitive states.

The limit of their approach is that the devices using to collect the data were mostly focused on collecting emotions and the data analysis was focused on finding correlation between emotions and progress, which was the core of the study. Monitoring the state of the mind in depth was not their purpose so that analysis was not precise, and was also limited because: (i) the assessment of emotions was performed subjectively by the participants; (ii) a single channel electroencephalogram (EEG) device was used, which may result in an error of up to fifty percent \cite{maskeliunas2016consumer}. 

Apel with colleagues study the work of the brain using very accurate techniques, like the functional magnetic resonance imaging (fMRI). 
They detected activation specific Broadmann-areas during code comprehension \cite{siegmund2014understanding}.
In their follow up work they investigated the difference between bottom-up program comprehension and comprehension with semantic cues in terms of brain areas involved \cite{DBLP:conf/sigsoft/SiegmundPPAHKBB17}.
A group led by Heui Seok Lim uses a full EEG device, like the one proposed in this research. However, they focus mostly on exploring how the mind of developers evolved from novice to experts in program comprehension tasks, and therefore have a completely different focus than ours \cite{lee2016comparing}.


\section{\uppercase{Problem statement}}
\label{problem}
Constant stress leads to physiological disadaptation with increased fatigability and to burn-out syndrome with decreased motivation for work, and thus inability to perform such important tasks. The main question is therefore, how it is possible to develop software when the stress is there and cannot be eliminated, but needs to be somehow mitigated to ensure high quality work. In other terms, the research problem is to find and study the best operating conditions for software that is developed under major time and psychological pressure for the developers like, for instance, when a remotely operated spacecraft is moving to an undesired location due to an error in the software, or a controller of a pipeline is wrongly operating causing leakages of gas.

Moreover, we will consider the following two sub-problems:
(i) when the stress occurs in a limited period of time, like when there is the need to fix a single safety-critical error within one working day;
(ii) when the stress spans longer intervals, like when a safety critical condition arises on a whole system, so that multiple days or weeks of work are required. 

These two scenarios require different approaches, since on the first case, very intense working patterns can be adopted, taking into account that a compensation might occur in the immediate future after the stress has occurred, while in the second there should be a pattern of work ensuring the ability to maintain the quality and productivity of a team for a longer period of time.
In detail, specific research questions that can be addressed are:\\
 \textbf{RQ1:} what is the effect of stress induced by the work-conditions on the mind of software developers and engineers and the implication of this stress on the quality of the software systems being produced,\\
\textbf{RQ2:} what mind states can be observed in software developers and engineers during stressful working conditions that are associated with either low or high quality and productivity, and then, specifically: {\bf (a)}  what are the typical working conditions when the stress results in low quality work or in loss of productivity, and how these condition can be mapped on mind states of developers, and {\bf (b)} what are the detailed software development processes or key individual processes and products patterns and practices that are observed to correlate with high quality and the productivity during critical circumstances and what are the associated mind states of software developers and engineers,\\
\textbf{RQ3:} what software development processes or key individual processes and products patterns and practices, or other actions can be elaborated to recreate in software developers and engineers the mind states that are typically associated with high quality and productivity, and how they can be elaborated within specific software development environments,\\
\textbf{RQ4:} what are the quantitative effects of the application of such processes and key individual processes and products patterns and practices in terms of productivity and quality of the generated software as functions of the condition of their use (a mapping between a working context and a problem to face).

The research questions require a lot of practical effort and preparation of specific environments to answer. In the next section we propose an approach to develop such an environment based on neuroimaging techniques, non-invasive software measurement tools and methods for evaluation of individual processes and products patterns in software engineering.

\section{\uppercase{Proposed approach}}
\label{approach}
Despite the recent trend of using neuroimaging techniques such as fMRI to understand the mind of developers, most work has been focused on general understanding of developers’ mind in the general context. As mentioned above such work has contributed to the understanding of developers’ mind, but it is often not clear how those general understandings can be applied to concrete real-world problems in software industry. In contrast, our approach focuses on a specific and critical context, that is, software development under stress-inducing circumstances. Understanding of developers’ mind in this specific context is not only scientifically novel, but also can make practical impact on software development practices. 

While we exploit emerging neuroimaging techniques (in particular, multi-channel EEG), these new techniques do not directly show which development methodologies and practices lead to better performance. Best development methodologies and practices can be identified only after considering not only neroimages but also other numerous factors related to developers and the artifacts created by the developers. Thus, the approach should support understanding not only mental effects of various software development practices on the performance of developers working in stressful environments. The key feature of the approach is systematic investigation of the practices from most relevant points. 

\subsection{Outline of the research agenda}
In this position paper we describe our agenda for the future research in the selected direction. At the first step we will select a family of software development processes for stressful circumstances. Further, we will collect key individual processes and products patterns and practices that are particularly useful when the critical circumstances arise. Next, we develop a framework for quantitative evaluation of such processes and key individual processes and products patterns and practices in terms of: {\bf (a) } the conditions when best to use them (working context -- problem to face; development environment; kind of software being developed), and {\bf (b) } the results in terms of productivity and quality of the generated software.

Finally, it is necessary to develop a set of tools that can help software engineers practice software development processes appropriate for a given context. For example, when software engineers work in an emergency situation, they will be able to accomplish their work more effectively and efficiently, with the help of the provided tools. Moreover, a system of integrated tools based on existing physical devices and software components and supplemented by a suitable integration layer and additional analysis techniques to collect the experimental data and to analyze it, to produce the results mentioned above.

We propose not only a systematic adoption of non-invasive measurement techniques (including analysis of processes and of products - code repositories, issue tracking data, budgeting information), but we couple it on one side with more standard data collected via surveys and on the other to biophysical data collected through suitable wearable devices, like wearable EEG, eye tracking devices, etc. This is possible by our partnership with key software development organizations which produce software for safety and business critical applications. We will be able to collect data from a uniquely large set of environments. 

Various experimentation are naturally fit into the proposed research agenda and could be used by other researchers as the cornerstone for subsequent analyses and also for identifying additional possible interpretations and for proposing other processes and product patterns and practices. 

\subsection{Background research}
The proposed agenda will be implemented along the following lines: {\bf (a) }data collection, {\bf (b) } data analysis and model construction, and {\bf (c) } model validation and refinement.

The first line refers mostly to the data collection. The work in this area will start from the idea of non-invasive data collection recently revised and actualized with the system Innometrics, whose most recent description will be presented at the 33rd ACM Symposium on Applied Computing (SAC 2018) \cite{Bykov-SAC18}. 
The data collection at companies will also be performed through suitable questionnaires and surveys, using the best standards in the field as can be applied in software engineering, as described in \cite{pedrycz2011model,ivanov2016assessing}. 
Additional data will be collected from full capacity EEG devices, one instance of which is already in use for feasibility studies at Innopolis University using the on one side the recent experience collected in software engineering \cite{lee2016comparing} 
and on the other the decades long competence presence in neurosciences.

The second line refers to data analysis and model construction. As mentioned, the data will be analyzed using statistics and machine learning. Statistics will be the standard statistical tools, as described in \cite{moser2007empirical}, 
more advanced regression techniques, as described in \cite{di2013pair}, 
approaches coming from reliability growth models \cite{ivanov2016assessing}. 
In terms of machine learning, we propose to use simple models based on logistic regression, support vector machine and random indexes \cite{fronza2013failure}, 
techniques based on neural networks, 
fuzzy logic, 
granular computing \cite{pedrycz2015data}, 
etc. 
For generalization purposes, statistical meta-analysis will be adopted \cite{snezana2012meta}. 

The third line refers to model validation and refinement. Substantially, we will build a suitable experimental design and, around it, we will run our repeated experimentation with the goal of ensuring validity and generalizable models, along the lines of the work done in \cite{succi2001analysis,succi2003practical,succi2003investigation,paulson2004empirical,DBLP:journals/ijitwe/RossiSSS06,janes2006identification,rossi2012adoption,janes2013managing,coman2014cooperation} and \cite{russo2015mining}. 

\subsection{Infrastructure}

Our infrastructure mainly consists of a non-invasive metrics collection system integrated with hardware devices collecting biometrics data. Note that our metrics collection system will collect various metrics encompassing metrics related to biometrics data of developers, metrics about developer activities performed during software development, metrics about software development process, and metrics about software artifacts. Our metrics collection system will also include software packages for statistical analysis that can be used to analyze collected data. 

In fact, the metrics collection system opens a new door to research on various topics for which it is essential to collect credible data about developers and software artifacts they develop.
Regarding our infrastructure, we plan apply the system in two major studies: {\bf (1) } a holistic metrics collection system that can collect metrics related to biometrics data of developers, metrics about developer activities,  software development process, and software artifacts, and {\bf (2) } an initial evaluation of using our holistic metrics collection system for study of software development under stress-inducing circumstances.

\subsection{User studies and experiments with students and developers}

With user study with industry partners, we expect to identify common practices exercised by developers to deal with stresses in their workplaces. We plan to report new findings we expect to find to major software engineering conferences and workshops. Note that such findings will not only enable our research, but also can help other researchers investigate the issues on stresses of developers. 

Based on the findings we obtain from the user study, we plan to investigate the actual effects of the practices exercised in the field to deal with stresses of developers. We also investigate the effects of these practices on performances and productivity of developers. For the sake of feasibility, we first plan to experiment with students of Innopolis University, and, based on the results we obtain, we also plan to perform similar experiments in industry partners. 

\section{\uppercase{Conclusion}}
\label{S.Conclusions}

Developing software systems is a knowledge intensive task, and as such is heavily influenced by the state of mind of developers. It has therefore historically been claimed that software has to be developed in a quiet and relaxed environment. However, this is hardly the case. Software is often produced under significant time constraint. Sometimes it even happens that patches for safety critical systems have to be released because one of such system is malfunctioning or not working at all with severe and even fatal consequences for its intended users. Notable examples for this include the aircraft and transportation industry and the overall energy industry. 

The main idea presented in this paper is to understand the effects of various software development practices on the performance of developers working in stressful environments, and identify the best operating conditions for software developed under stressful conditions. We discussed the possible research agenda and provide our view on its implementation with the state of the art technologies and approaches. 

\section*{\uppercase{Acknowledgements}}
We thank Innopolis University for generously funding this research.


\bibliographystyle{apalike}
{\small
\bibliography{ENASE.SoftwareUnderStressProposal}}

\begin{thebibliography}{}

\bibitem[Ali et~al., 2014]{ali2014systematic}
Ali, N.~B., Petersen, K., and Wohlin, C. (2014).
\newblock A systematic literature review on the industrial use of software
  process simulation.
\newblock {\em Journal of Systems and Software}, 97:65--85.

\bibitem[Amabile, 1996]{Amabile-Harvard96}
Amabile, T.~M. (1996).
\newblock Creativity and innovation in organizations.
\newblock {\em Harvard Business School Background Note}, pages 396--239.

\bibitem[Baas et~al., 2008]{Baas-Psychological08}
Baas, M., De~Dreu, C., and Nijstad, B. (2008).
\newblock A meta-analysis of 25 years of mood-creativity research: Hedonic
  tone, activation, or regulatory focus?
\newblock {\em Psychological Bulletin}, 134:779--806.

\bibitem[Barsade and Gibson, 2007]{barsade2007does}
Barsade, S.~G. and Gibson, D.~E. (2007).
\newblock Why does affect matter in organizations?
\newblock {\em The Academy of Management Perspectives}, 21(1):36--59.

\bibitem[Brand et~al., 2007]{brand2007we}
Brand, S., Reimer, T., and Opwis, K. (2007).
\newblock How do we learn in a negative mood? effects of a negative mood on
  transfer and learning.
\newblock {\em Learning and instruction}, 17(1):1--16.

\bibitem[Bykov et~al., 2018]{Bykov-SAC18}
Bykov, A., Ivanov, V., Rogers, A., Shunevich, A., Sillitti, A., Succi, G.,
  Tormasov, A., Yi, J., Zabirov, A., and Zaplatnikov, D. (2018).
\newblock A new architecture and implementation strategy for non-invasive
  software measurement systems.
\newblock In {\em Proceedings of the 33rd ACM/SIGAPP Symposium On Applied
  Computing (SAC 2018)}. ACM.
\newblock To appear.

\bibitem[Campanelli and Parreiras, 2015]{campanelli2015agile}
Campanelli, A.~S. and Parreiras, F.~S. (2015).
\newblock Agile methods tailoring--a systematic literature review.
\newblock {\em Journal of Systems and Software}, 110:85--100.

\bibitem[Cockburn and Highsmith, 2001]{cockburn2001agile}
Cockburn, A. and Highsmith, J. (2001).
\newblock Agile software development, the people factor.
\newblock {\em Computer}, 34(11):131--133.

\bibitem[Coman et~al., 2014]{coman2014cooperation}
Coman, I.~D., Robillard, P.~N., Sillitti, A., and Succi, G. (2014).
\newblock Cooperation, collaboration and pair-programming: Field studies on
  backup behavior.
\newblock {\em Journal of Systems and Software}, 91:124--134.

\bibitem[Corral et~al., 2013a]{Succi:C244.2013}
Corral, L., Georgiev, A.~B., Sillitti, A., and Succi, G. (2013a).
\newblock {A method for characterizing energy consumption in Android
  smartphones}.
\newblock In {\em Green and Sustainable Software (GREENS 2013), 2nd
  International Workshop on}, pages 38--45. IEEE.

\bibitem[Corral et~al., 2013b]{Succi:C245.2013}
Corral, L., Sillitti, A., and Succi, G. (2013b).
\newblock {Software development processes for mobile systems: Is agile really
  taking over the business?}
\newblock In {\em {Engineering of Mobile-Enabled Systems (MOBS), 2013 1st
  International Workshop on the}}, pages 19--24.

\bibitem[Corral et~al., 2011]{Succi:C236.2011}
Corral, L., Sillitti, A., Succi, G., Garibbo, A., and Ramella, P. (2011).
\newblock {Evolution of Mobile Software Development from Platform-Specific to
  Web-Based Multiplatform Paradigm}.
\newblock In {\em Proceedings of the 10th SIGPLAN Symposium on New Ideas, New
  Paradigms, and Reflections on Programming and Software}, Onward! 2011, pages
  181--183, New York, NY, USA. ACM.

\bibitem[Corral et~al., 2012]{DBLP:conf/tools/CorralSSSV12}
Corral, L., Sillitti, A., Succi, G., Strumpflohner, J., and Vlasenko, J.
  (2012).
\newblock Droidsense: {A} mobile tool to analyze software development processes
  by measuring team proximity.
\newblock In {\em Proceedings of the the 50th International Conference on
  Objects, Models, Components, Patterns (TOOLS Europe 2012)}, pages 17--33.

\bibitem[Davis, 2009]{davis2009understanding}
Davis, M.~A. (2009).
\newblock Understanding the relationship between mood and creativity: A
  meta-analysis.
\newblock {\em Organizational behavior and human decision processes},
  108(1):25--38.

\bibitem[Denning, 2012]{denning2012moods}
Denning, P.~J. (2012).
\newblock Moods.
\newblock {\em Communications of the ACM}, 55(12):33--35.

\bibitem[di~Bella et~al., 2013]{di2013pair}
di~Bella, E., Fronza, I., Phaphoom, N., Sillitti, A., Succi, G., and Vlasenko,
  J. (2013).
\newblock Pair programming and software defects--a large, industrial case
  study.
\newblock {\em IEEE Transactions on Software Engineering}, 39(7):930--953.

\bibitem[Di~Bella et~al., 2013]{Succi:J91.2013}
Di~Bella, E., Sillitti, A., and Succi, G. (2013).
\newblock A multivariate classification of open source developers.
\newblock {\em Information Sciences}, 221:72--83.

\bibitem[Djokic et~al., 2012]{snezana2012meta}
Djokic, S., Succi, G., Pedrycz, W., and Mintchev, M. (2012).
\newblock Meta analysis--a method of combining empirical results and its
  application in object-oriented software systems.
\newblock In {\em OOIS 2001: 7th International Conference on Object-Oriented
  Information Systems 27--29 August 2001, Calgary, Canada}, pages 103--112.
  Springer Science \& Business Media.

\bibitem[Feldt et~al., 2010]{feldt2010links}
Feldt, R., Angelis, L., Torkar, R., and Samuelsson, M. (2010).
\newblock Links between the personalities, views and attitudes of software
  engineers.
\newblock {\em Information and Software Technology}, 52(6):611--624.

\bibitem[Fischer, 1987]{fischer1987cognitive}
Fischer, G. (1987).
\newblock Cognitive view of reuse and redesign.
\newblock {\em IEEE Software}, 4(4):60.

\bibitem[Fronza et~al., 2009]{Succi:C217.2009}
Fronza, I., Sillitti, A., and Succi, G. (2009).
\newblock {An Interpretation of the Results of the Analysis of Pair Programming
  During Novices Integration in a Team}.
\newblock In {\em Proceedings of the 2009 3rd International Symposium on
  Empirical Software Engineering and Measurement}, ESEM '09, pages 225--235.
  IEEE Computer Society.

\bibitem[Fronza et~al., 2013]{fronza2013failure}
Fronza, I., Sillitti, A., Succi, G., Terho, M., and Vlasenko, J. (2013).
\newblock Failure prediction based on log files using random indexing and
  support vector machines.
\newblock {\em Journal of Systems and Software}, 86(1):2--11.

\bibitem[Glass et~al., 1992]{glass1992software}
Glass, R.~L., Vessey, I., and Conger, S.~A. (1992).
\newblock Software tasks: Intellectual or clerical?
\newblock {\em Information \& Management}, 23(4):183--191.

\bibitem[Graziotin et~al., 2013]{graziotin2013happy}
Graziotin, D., Wang, X., and Abrahamsson, P. (2013).
\newblock Are happy developers more productive?
\newblock In {\em Proceedings of the 2013 International Conference on Product
  Focused Software Process Improvement}, pages 50--64. Springer.

\bibitem[Graziotin et~al., 2014]{graziotin2014happy}
Graziotin, D., Wang, X., and Abrahamsson, P. (2014).
\newblock Happy software developers solve problems better: psychological
  measurements in empirical software engineering.
\newblock {\em PeerJ}, 2:e289.

\bibitem[Ivanov et~al., 2016]{ivanov2016assessing}
Ivanov, V., Mazzara, M., Pedrycz, W., Sillitti, A., and Succi, G. (2016).
\newblock Assessing the process of an eastern european software sme using
  systemic analysis, gqm, and reliability growth models-a case study.
\newblock In {\em Proceedings of the 2016 IEEE/ACM International Conference on
  Software Engineering Companion (ICSE-C)}, pages 251--259. IEEE.

\bibitem[Janes et~al., 2013]{janes2013managing}
Janes, A., Remencius, T., Sillitti, A., and Succi, G. (2013).
\newblock Managing changes in requirements: an empirical investigation.
\newblock {\em Journal of software: evolution and process}, 25(12):1273--1283.

\bibitem[Janes et~al., 2006]{janes2006identification}
Janes, A., Scotto, M., Pedrycz, W., Russo, B., Stefanovic, M., and Succi, G.
  (2006).
\newblock Identification of defect-prone classes in telecommunication software
  systems using design metrics.
\newblock {\em Information sciences}, 176(24):3711--3734.

\bibitem[Janes and Succi, 2012]{Succi:c244.2012}
Janes, A.~A. and Succi, G. (2012).
\newblock The dark side of agile software development.
\newblock In {\em Proceedings of the ACM International Symposium on New Ideas,
  New Paradigms, and Reflections on Programming and Software}, Onward! 2012,
  pages 215--228, New York, NY, USA. ACM.

\bibitem[Khan et~al., 2017]{khan2017systematic}
Khan, A.~A., Keung, J., Niazi, M., Hussain, S., and Zhang, H. (2017).
\newblock Systematic literature reviews of software process improvement: A
  tertiary study.
\newblock In {\em Proceedings of the 2017 European Conference on Software
  Process Improvement}, pages 177--190. Springer.

\bibitem[Khan et~al., 2011]{khan2011moods}
Khan, I.~A., Brinkman, W.-P., and Hierons, R.~M. (2011).
\newblock Do moods affect programmers' debug performance?
\newblock {\em Cognition, Technology \& Work}, 13(4):245--258.

\bibitem[Kivi et~al., 2000]{Succi:C73.2000}
Kivi, J., Haydon, D., Hayes, J., Schneider, R., and Succi, G. (2000).
\newblock Extreme programming: a university team design experience.
\newblock In {\em 2000 Canadian Conference on Electrical and Computer
  Engineering. Conference Proceedings. Navigating to a New Era (Cat.
  No.00TH8492)}, volume~2, pages 816--820 vol.2.

\bibitem[Knobelsdorf and Romeike, 2008]{Knobelsdorf:2008:CPC:1597849.1384347}
Knobelsdorf, M. and Romeike, R. (2008).
\newblock Creativity as a pathway to computer science.
\newblock {\em SIGCSE Bulletin}, 40(3):286--290.

\bibitem[Kov{\'a}cs et~al., 2004]{Succi:C144.2004}
Kov{\'a}cs, G.~L., Drozdik, S., Zuliani, P., and Succi, G. (2004).
\newblock {Open Source Software for the Public Administration}.
\newblock In {\em Proceedings of the 6th International Workshop on Computer
  Science and Information Technologies}.

\bibitem[Lee et~al., 2016]{lee2016comparing}
Lee, S., Matteson, A., Hooshyar, D., Kim, S., Jung, J., Nam, G., and Lim, H.
  (2016).
\newblock Comparing programming language comprehension between novice and
  expert programmers using eeg analysis.
\newblock In {\em Proceedings of the IEEE 16th International Conference on
  Bioinformatics and Bioengineering (BIBE)}, pages 350--355.

\bibitem[Lewis et~al., 2011]{lewis2011affective}
Lewis, S., Dontcheva, M., and Gerber, E. (2011).
\newblock Affective computational priming and creativity.
\newblock In {\em Proceedings of the 2011 SIGCHI Conference on Human Factors in
  Computing Systems}, pages 735--744.

\bibitem[Lyubomirsky et~al., 2005]{lyubomirsky2005benefits}
Lyubomirsky, S., King, L., and Diener, E. (2005).
\newblock The benefits of frequent positive affect: Does happiness lead to
  success?
\newblock {\em Psychological Bulletin}, 131(6):803--855.

\bibitem[Marino and Succi, 1989]{Succi:C2.1989}
Marino, G. and Succi, G. (1989).
\newblock {Data Structures for Parallel Execution of Functional Languages}.
\newblock In {\em Proceedings of the Parallel Architectures and Languages
  Europe, Volume II: Parallel Languages}, PARLE '89, pages 346--356.
  Springer-Verlag.

\bibitem[Maskeliunas et~al., 2016]{maskeliunas2016consumer}
Maskeliunas, R., Damasevicius, R., Martisius, I., and Vasiljevas, M. (2016).
\newblock Consumer-grade eeg devices: are they usable for control tasks?
\newblock {\em PeerJ}, 4:e1746.

\bibitem[Moser et~al., 2007]{moser2007empirical}
Moser, R., Russo, B., and Succi, G. (2007).
\newblock Empirical analysis on the correlation between gcc compiler warnings
  and revision numbers of source files in five industrial software projects.
\newblock {\em Empirical Software Engineering}, 12(3):295--310.

\bibitem[M{\"u}ller and Fritz, 2015]{muller2015stuck}
M{\"u}ller, S.~C. and Fritz, T. (2015).
\newblock Stuck and frustrated or in flow and happy: Sensing developers'
  emotions and progress.
\newblock In {\em International Conference on the IEEE/ACM 37th IEEE Software
  Engineering (ICSE)}, volume~1, pages 688--699. IEEE.

\bibitem[Murgia et~al., 2014]{murgia2014developers}
Murgia, A., Tourani, P., Adams, B., and Ortu, M. (2014).
\newblock Do developers feel emotions? an exploratory analysis of emotions in
  software artifacts.
\newblock In {\em Proceedings of the 11th working conference on mining software
  repositories}, pages 262--271.

\bibitem[Paulson et~al., 2004]{paulson2004empirical}
Paulson, J.~W., Succi, G., and Eberlein, A. (2004).
\newblock An empirical study of open-source and closed-source software
  products.
\newblock {\em IEEE Transactions on Software Engineering}, 30(4):246--256.

\bibitem[Pedrycz et~al., 2011]{pedrycz2011model}
Pedrycz, W., Russo, B., and Succi, G. (2011).
\newblock A model of job satisfaction for collaborative development processes.
\newblock {\em Journal of Systems and Software}, 84(5):739--752.

\bibitem[Pedrycz et~al., 2015a]{Succi:J100.2015}
Pedrycz, W., Succi, G., Sillitti, A., and Iljazi, J. (2015a).
\newblock Data description: {A} general framework of information granules.
\newblock {\em Knowledge-Based Systems}, 80:98--108.

\bibitem[Pedrycz et~al., 2015b]{pedrycz2015data}
Pedrycz, W., Succi, G., Sillitti, A., and Iljazi, J. (2015b).
\newblock Data description: a general framework of information granules.
\newblock {\em Knowledge-Based Systems}, 80:98--108.

\bibitem[Petrinja et~al., 2010]{Succi:C224.2010}
Petrinja, E., Sillitti, A., and Succi, G. (2010).
\newblock {Comparing OpenBRR, QSOS, and OMM assessment models}.
\newblock In {\em {Open Source Software: New Horizons - Proceedings of the 6th
  International IFIP WG 2.13 Conference on Open Source Systems, OSS 2010}},
  pages 224--238, Notre Dame, IN, USA. Springer, Heidelberg.

\bibitem[Rossi et~al., 2012a]{Succi:J83.2012}
Rossi, B., Russo, B., and Succi, G. (2012a).
\newblock Adoption of free/libre open source software in public organizations:
  factors of impact.
\newblock {\em Information Technology \& People}, 25(2):156--187.

\bibitem[Rossi et~al., 2012b]{rossi2012adoption}
Rossi, B., Russo, B., and Succi, G. (2012b).
\newblock Adoption of free/libre open source software in public organizations:
  factors of impact.
\newblock {\em Information Technology \& People}, 25(2):156--187.

\bibitem[Rossi et~al., 2006]{DBLP:journals/ijitwe/RossiSSS06}
Rossi, B., Scotto, M., Sillitti, A., and Succi, G. (2006).
\newblock An empirical study on the migration to openoffice.org in a public
  administration.
\newblock {\em {International Journal of Information Technology and Web
  Engineering}}, 1(3):64--80.

\bibitem[Rowan, 1957]{rowan1957psychological}
Rowan, T.~C. (1957).
\newblock Psychological tests and selection of computer programmers.
\newblock {\em Journal of the ACM (JACM)}, 4(3):348--353.

\bibitem[Russo et~al., 2015]{russo2015mining}
Russo, B., Succi, G., and Pedrycz, W. (2015).
\newblock Mining system logs to learn error predictors: a case study of a
  telemetry system.
\newblock {\em Empirical Software Engineering}, 20(4):879--927.

\bibitem[Shaw, 2004]{shaw2004emotions}
Shaw, T. (2004).
\newblock The emotions of systems developers: an empirical study of affective
  events theory.
\newblock In {\em Proceedings of the 2004 SIGMIS conference on Computer
  personnel research: Careers, culture, and ethics in a networked environment},
  pages 124--126. ACM.

\bibitem[Siegmund et~al., 2014]{siegmund2014understanding}
Siegmund, J., K{\"a}stner, C., Apel, S., Parnin, C., Bethmann, A., Leich, T.,
  Saake, G., and Brechmann, A. (2014).
\newblock Understanding understanding source code with functional magnetic
  resonance imaging.
\newblock In {\em Proceedings of the 36th International Conference on Software
  Engineering}, pages 378--389. ACM.

\bibitem[Siegmund et~al., 2017]{DBLP:conf/sigsoft/SiegmundPPAHKBB17}
Siegmund, J., Peitek, N., Parnin, C., Apel, S., Hofmeister, J., K{\"{a}}stner,
  C., Begel, A., Bethmann, A., and Brechmann, A. (2017).
\newblock Measuring neural efficiency of program comprehension.
\newblock In {\em Proceedings of the 2017 11th Joint Meeting on Foundations of
  Software Engineering, {ESEC/FSE}}, pages 140--150.

\bibitem[Sillitti et~al., 2004]{Succi:J60.2004}
Sillitti, A., Janes, A., Succi, G., and Vernazza, T. (2004).
\newblock Measures for mobile users: an architecture.
\newblock {\em Journal of Systems Architecture}, 50(7):393--405.

\bibitem[Sillitti et~al., 2012]{sillitti2012understanding}
Sillitti, A., Succi, G., and Vlasenko, J. (2012).
\newblock Understanding the impact of pair programming on developers attention:
  a case study on a large industrial experimentation.
\newblock In {\em Proceedings of the 34th International Conference on Software
  Engineering (ICSE)}, pages 1094--1101.

\bibitem[Succi et~al., 2001]{succi2001analysis}
Succi, G., Benedicenti, L., and Vernazza, T. (2001).
\newblock Analysis of the effects of software reuse on customer satisfaction in
  an rpg environment.
\newblock {\em IEEE Transactions on Software Engineering}, 27(5):473--479.

\bibitem[Succi et~al., 2002]{succi2002preliminary}
Succi, G., Pedrycz, W., Marchesi, M., and Williams, L. (2002).
\newblock Preliminary analysis of the effects of pair programming on job
  satisfaction.
\newblock In {\em Proceedings of the 3rd International Conference on Extreme
  Programming (XP)}, pages 212--215.

\bibitem[Succi et~al., 2003a]{succi2003practical}
Succi, G., Pedrycz, W., Stefanovic, M., and Miller, J. (2003a).
\newblock Practical assessment of the models for identification of defect-prone
  classes in object-oriented commercial systems using design metrics.
\newblock {\em Journal of systems and software}, 65(1):1--12.

\bibitem[Succi et~al., 2003b]{succi2003investigation}
Succi, G., Pedrycz, W., Stefanovic, M., and Russo, B. (2003b).
\newblock An investigation on the occurrence of service requests in commercial
  software applications.
\newblock {\em Empirical Software Engineering}, 8(2):197--215.

\bibitem[Sulayman and Mendes, 2009]{sulayman2009systematic}
Sulayman, M. and Mendes, E. (2009).
\newblock A systematic literature review of software process improvement in
  small and medium web companies.
\newblock {\em Advances in software engineering}, pages 1--8.

\bibitem[Unterkalmsteiner et~al., 2012]{unterkalmsteiner2012evaluation}
Unterkalmsteiner, M., Gorschek, T., Islam, A.~M., Cheng, C.~K., Permadi, R.~B.,
  and Feldt, R. (2012).
\newblock Evaluation and measurement of software process improvement--a
  systematic literature review.
\newblock {\em IEEE Transactions on Software Engineering}, 38(2):398--424.

\bibitem[Valerio et~al., 1997]{Succi:J20.1997}
Valerio, A., Succi, G., and Fenaroli, M. (1997).
\newblock Domain analysis and framework-based software development.
\newblock {\em SIGAPP Appl. Comput. Rev.}, 5(2):4--15.

\bibitem[Vernazza et~al., 2000]{Succi:C82.2000}
Vernazza, T., Granatella, G., Succi, G., Benedicenti, L., and Mintchev, M.
  (2000).
\newblock {Defining Metrics for Software Components}.
\newblock In {\em Proceedings of the World Multiconference on Systemics,
  Cybernetics and Informatics}, volume~XI, pages 16--23.

\bibitem[Williams and Cockburn, 2003]{DBLP:journals/computer/WilliamsC03}
Williams, L.~A. and Cockburn, A. (2003).
\newblock Guest editors' introduction: Agile software development: It's about
  feedback and change.
\newblock {\em {IEEE} Computer}, 36(6):39--43.

\bibitem[Z{\"u}ger and Fritz, 2015]{zuger2015interruptibility}
Z{\"u}ger, M. and Fritz, T. (2015).
\newblock Interruptibility of software developers and its prediction using
  psycho-physiological sensors.
\newblock In {\em Proceedings of the 33rd Annual ACM Conference on Human
  Factors in Computing Systems}, pages 2981--2990. ACM.

\end{thebibliography}

\end{document}